\documentclass[a4paper,fleqn,usenatbib]{mnras}
\usepackage[T1]{fontenc}
\usepackage{ae,aecompl}
\usepackage{graphicx}	
\usepackage{amsmath}	
\usepackage{amssymb}	
\usepackage{float}
\usepackage{color}

\usepackage[draft]{hyperref}	
\hypersetup{colorlinks=true,linkcolor=blue,citecolor=blue,filecolor=blue,urlcolor=blue}

\title{Improving the convergence properties of the moving-mesh code
  AREPO} \author[R.~Pakmor et al.]  {R\"udiger~Pakmor$^1$\thanks{E-mail: ruediger.pakmor@h-its.org},
  Volker~Springel$^{1,2}$, Andreas~Bauer$^{1}$, Philip~Mocz$^{3}$,
  Diego~J.~Munoz$^{4}$,
  \newauthor Sebastian~T.~Ohlmann$^{5,1}$, Kevin~Schaal$^{1,2}$, Chenchong~Zhu$^{6}$ \vspace*{0.2cm}\\
  $^1$Heidelberger Institut f\"{u}r Theoretische Studien,
  Schloss-Wolfsbrunnenweg 35, 69118 Heidelberg, Germany\\
  $^2$Zentrum f\"ur Astronomie der Universit\"at Heidelberg,
  Astronomisches Recheninstitut, M\"{o}nchhofstr. 12-14, 69120
  Heidelberg, Germany \\
  $^3$Harvard-Smithsonian Center for Astrophysics, 60 Garden Street,
  Cambridge, MA 02138, USA \\
  $^4$Center for Space Research, Department of Astronomy, Cornell University, Ithaca, NY 14853, USA\\
  $^5$Institut f\"{u}r Theoretische Physik und Astrophysik,
  Universit\"{a}t W\"{u}rzburg,
  Am Hubland, D-97074 W\"{u}rzburg, Germany\\
  $^6$Department of Astronomy \& Astrophysics, University of Toronto, 50 St. George Street, Toronto, Ontario, Canada, M5S 3H4\\
}
  
\pubyear{2015}

\begin{document}

\label{firstpage}
\pagerange{\pageref{firstpage}--\pageref{lastpage}}

\maketitle

\begin{abstract}
  Accurate numerical solutions of the equations of hydrodynamics play
  an ever more important role in many fields of astrophysics.  In this
  work, we reinvestigate the accuracy of the moving-mesh code
  \textsc{Arepo} and show how its convergence order can be improved
  for general problems. In particular, we clarify that for certain
  problems \textsc{Arepo} only reaches first-order convergence for its
  original formulation. This can be rectified by simple modifications
  we propose to the time integration scheme and the spatial gradient
  estimates of the code, both improving the accuracy of the code. We
  demonstrate that the new implementation is indeed second-order
  accurate under the $L^1$ norm, and in particular substantially
  improves conservation of angular momentum. Interestingly, whereas
  these improvements can significantly change the results of smooth
  test problems, we also find that cosmological simulations of galaxy
  formation are unaffected, demonstrating that the numerical errors
  eliminated by the new formulation do not impact these
  simulations. In contrast, simulations of binary stars followed over
  a large number of orbital times are strongly affected, as here it is
  particularly crucial to avoid a long-term build up of errors in
  angular momentum conservation.
\end{abstract}

\begin{keywords}
  methods: numerical, hydrodynamics, galaxy: formation
\end{keywords}

\section{Introduction}

Gravity and hydrodynamics provide the foundations for describing
almost all phenomena in astrophysics. Often, the dynamics and geometry
are sufficiently complicated that analytic solutions, especially for
hydrodynamics, cannot be obtained. These limitations can be overcome
by numerical approaches, which have developed over the recent decades
into powerful tools for studying complex hydrodynamical flow
problems. It is however an ongoing and persistent challenge to derive
ever better discretization schemes that reduce numerical
discretization errors to a minimum, while at the same time offering
high adaptivity to different time- and length-scales and a high degree
of parallel scalability. Astrophysical code development is therefore
best viewed as an iterative effort which never is fully complete and
finished. Rather, new generations of numerical `instruments'
(i.e.~codes) should ideally improve their accuracy and speed, somewhat
similar in spirit to the advances regularly realized in observational
instrumentation.

Traditionally, stationary mesh codes with or without adaptive mesh
refinement \citep[e.g.][]{FLASH,Teyssier2002,Castro,Athena,ENZO} and
smoothed particle hydrodynamics (SPH) codes
\citep[e.g.][]{Wadsley2004,Springel2005b} have dominated astrophysical
fluid dynamics. Moving meshes and hybrid techniques combining
Eulerian and Lagrangian meshes have been applied only rarely in astrophysics
and related fields \citep[e.g.][]{hirt1974,gnedin1995a,pen1998a}.
However, in particular in astrophysics they have not seen widespread use so far.

Recently, moving-mesh techniques based on a Voronoi mesh have been 
proposed \citep{Arepo,Duffell2011a}, which offer a quasi-Lagrangian description
that inherits some of the advantages of SPH while retaining the
accuracy of a Eulerian mesh-based description. These new methods have
matured into interesting alternative codes that have already been
widely applied in cosmological simulations of structure formation
\citep[e.g.][]{Vogelsberger2012, Sijacki2012, Marinacci2013,
  Vogelsberger2014Illustris, Genel2014}, as well as in first star
formation \citep[e.g.][]{Greif2011, Greif2012, Smith2014}, and to a
smaller extent in stellar astrophysics \citep{Pakmor2013SN} and in
problems related to planet formation \citep{Duffell2012, Munoz2014,
  Munoz2015}. It has become clear that it can be very worthwhile to
accept the technical complications introduced by a dynamic and
unstructured mesh given the benefits realized by combining most of the
advantages of fixed mesh codes (e.g. better convergence for smooth
flows, good shock capturing) and SPH techniques (e.g. intrinsic
adaptivity, lack of advection errors, small numerical diffusion).
 
A well-defined and rapid convergence rate is a particularly important
feature of mesh codes, which are usually constructed to be at least
second-order accurate for smooth flows, i.e.~in the absence of shocks
or other discontinuities. For SPH codes, in contrast, it is very hard
to demonstrate proper convergence in the first place
\citep{Zhu2015}. In practice, the convergence rate dictates how
rapidly a numerical solution improves as more computational effort is
invested, an aspect that is arguably even more important than the
absolute size of the error itself. For example, while SPH often
obtains a qualitatively correct result with acceptable error at low
resolution, its poor convergence rate does not allow it to reach
comparable accuracy to mesh codes at comparable numerical cost once
the resolution is high. In contrast state of the art mesh-based codes
are expected to show at least second order convergence.

Moving-meshes in general have been studied extensively 
\citep[see, e.g.][]{Thomas1979,guillard2000a}.
In this paper, we reexamine the accuracy and convergence rate of the
\textsc{Arepo} code \citep{Arepo}, which is at present the most widely
employed implementation of the moving-mesh technique in astrophysics.
We address weaknesses in the original version of this code and show how they can
be overcome to improve the overall accuracy of the scheme. This
directly affects the range of applicability of the code and is hence
important for further maturing the moving-mesh approach in general.

The paper is structured as follows. In Section~\ref{sec:current}, we
concisely review the essential parts of the implementation of
moving-mesh hydrodynamics in \textsc{Arepo} and analyse its
convergence. In Sections~\ref{sec:rk} and~\ref{sec:lsf}, we describe
improvements to the time integration scheme and the gradient estimate,
respectively. We apply the new implementation to test problems in
Section~\ref{sec:tests}, show results for realistic applications in
Section~\ref{sec:world}, and discuss the implications of our work in
Section~\ref{sec:conclusion}.
    
\section{Moving-mesh hydrodynamics in \textsc{Arepo}}
\label{sec:current}

The moving-mesh code \textsc{Arepo} solves the Euler equations using
the finite-volume approach on an unstructured Voronoi mesh that is
generated from a set of points \citep{Arepo}. At any given time during
the simulation, the mesh can be uniquely constructed given only the
positions of the mesh-generating points. Since the mesh-generating
points move, the geometry of the mesh and its connectivity change over
time.
 
The Euler equations can be written as a system of hyperbolic
conservation laws for conserved quantities $\textbf{U}$ and a flux
function $\textbf{F}(\textbf{U})$,
 \begin{equation}
    \frac{\partial \textbf{U}}{\partial t} + \nabla\cdot \textbf{F} = 0 .
 \end{equation}
For the standard Euler equations the conserved quantities are mass, momentum
and energy. They define a vector of conserved quantities and an associated flux function as
\begin{equation}
  \textbf{U} = \left( \begin{array}{c} \rho \\ \rho \textbf{v} \\ \rho e \end{array} \right) ,
   \ \ \ \ \ \ \
   \textbf{F}(\textbf{U}) = \left( \begin{array}{c} 
    \rho \textbf{v} \\
    \rho \textbf{v} \textbf{v}^T + P  \\
    \rho e \textbf{v} + P \textbf{v}
  \end{array} \right),
\end{equation}
where $\rho$, $\textbf{v}$, $P$, and $e$ are density, velocity,
pressure, and total specific energy of the fluid, respectively. The
latter is defined as the sum of the specific internal energy $u$ and
the specific kinetic energy $\frac{1}{2} \textbf{v}^2$, thus $e = u + \frac{1}{2} \textbf{v}^2$. The
system of equation is closed by an equation of state 
$P = \rho (\gamma - 1) e$ with the adiabatic index $\gamma$.
 
To solve the Euler equations, the state of the fluid is discretised using
the cells of the Voronoi mesh. To this end, averages of the conserved
quantities $\textbf{U}$ of the fluid are computed for each cell 
through integration over the finite volume of a cell $i$, 
 \begin{equation}
   \textbf{Q}_i = \int_{V_i} \textbf{U} \, \mathrm{d}V.
 \end{equation}
yielding a census of the conserved physical quantities in the cell.
This state is then evolved in time by
\begin{equation}
   \textbf{Q}_i^{n+1} = \textbf{Q}_i^{n} - \Delta t \sum_j A_{ij} \mathbf{\hat{F}}_{ij}^{n+1/2},
   \label{eqn:timeInt}
\end{equation}
where $A_{ij}$ is the oriented area of the face between cells $i$ and
$j$ and $\mathbf{\hat{F}}$ is a time-averaged approximation of the
true flux $\textbf{F}_{ij}$ across the interface between cells $i$ and
$j$.
 
The calculation of the fluxes $\mathbf{\hat{F}}$ is done in
the original version of \textsc{Arepo} using a MUSCL-Hancock scheme
\citep{vanLeer1984,Toro}, which has been shown to provide second-order
accuracy in time and space for all kinds of stationary meshes
\citep[see, e.g.][]{FLASH}. This method uses a slope-limited piece-wise linear
spatial reconstruction step in each cell and a first order
time extrapolation of the fluid states by half a timestep to obtain
the states on both sides of all interfaces. Finally, a Riemann solver
uses the states on both sides of an interface to compute the flux that
is exchanged during the timestep.
 
The extrapolation is carried out in primitive variables 
\begin{equation}
   \textbf{W} = \left( \begin{array}{c} \rho \\ \textbf{v} \\ P \end{array} \right),
\end{equation}
which are straightforwardly calculated from the conserved quantities
and the geometry of a cell with the equation of state, combined with
estimates for their local spatial gradients. By construction, they
represent the values at the center of mass of a cell. Then, the left and right
states of an interface are computed by a linear spatial extrapolation
from the center of mass $\mathbf{s}$ of a cell to the center
$\mathbf{f}$ of a cell face, and by a half-step prediction forward in
time (where $\Delta t$ is the full timestep) as
\begin{equation}
   \textbf{W}^{\prime}_{\mathrm{L,R}} = \textbf{W}_{\mathrm{L,R}} 
   + \left. \frac{\partial \textbf{W}}{\partial \textbf{r}} \right|_{\mathrm{L,R}} \left( \mathbf{f} - \mathbf{s}_\mathrm{L,R} \right)
   + \left. \frac{\partial \textbf{W}}{\partial \textbf{t}} \right|_{\mathrm{L,R}} \frac{\Delta t}{2}.
   \label{eqn:ep}
\end{equation}

The time derivatives of the primitive variables in
Eqn.~(\ref{eqn:ep}) are not calculated directly. Instead, the
Euler equations are used to express them in terms of the primitive
variables and their spatial derivatives only:
\begin{eqnarray}
   \frac{ \partial \rho }{ \partial t } &=& - \textbf{v} \nabla \rho - \rho \nabla \textbf{v},  \\
   \frac{ \partial \textbf{v} }{ \partial t } &=& - \frac{\nabla P}{ \rho } - \textbf{v} \nabla \textbf{v}^T , \\
   \frac{ \partial P }{ \partial t } &=& - \gamma P \nabla \textbf{v} - \textbf{v} \nabla P.
\end{eqnarray}
The extrapolation and computation of the fluxes over an interface are
all done in the rest frame of the (moving) interface, i.e. the interface
velocity is subtracted from the equations above. This allows
solutions that are manifestly Galilean-invariant, an important
conceptual advantage over traditional Eulerian methods where the
numerical truncation error grows with the fluid velocity.
 
In the original implementation of \textsc{Arepo}, the spatial gradients are 
calculated using an improved Green-Gauss estimator that makes use of
certain mathematical properties of the Voronoi mesh. Specifically, the
gradient estimate for a primitive variable $\phi$ in cell $i$ is given by
\begin{equation}
   \left< \nabla \phi \right>_i = \frac{1}{V_i} \sum_j A_{ij} \left( \left[ \phi_j - \phi_i \right] \frac{ \textbf{c}_{ij} }{ r_{ij} }
   - \frac{ \phi_i + \phi_j }{2} \frac{ \textbf{r}_{ij} }{ r_{ij} } \right),
\label{eqngradient}
\end{equation}
where
\begin{equation}
   \textbf{c}_{ij} = \frac{1}{A_{ij}} \int_{A_{ij}} \left( \textbf{r} - \frac{\textbf{r}_i + \textbf{r}_j}{2} \right) \mathrm{d}A
\end{equation}
is the center of mass of the face between $i$ and $j$, and
$\textbf{r}_{ij} = \textbf{r}_i - \textbf{r}_j$ is the difference
between the positions of the mesh-generation points of cells $i$ and
$j$, and $r_{ij} = \left| \textbf{r}_{ij} \right|$ is its length. The sum
runs over all cells that share an interface
with cell $i$.  Note that this estimate of the spatial gradient of a
primitive variable depends only on the values of the primitive
variable in the cell itself and its direct neighbours, and on the
local geometry of the cell. If the values of the primitive variables
sample an underlying smooth field at the {\em positions of the
  mesh-generating points}, then the gradient estimate is second-order
accurate for an {\em arbitrary mesh geometry}, i.e.~it reproduces a constant
gradient in the field to machine precision. However, the primitive variables
$\textbf{Q}$ used to compute the gradient estimates represent volume averaged
quantities rather than the value of the fluid at the mesh generating point of
a cell.

\begin{figure}
  \centering
  \includegraphics[width=\linewidth]{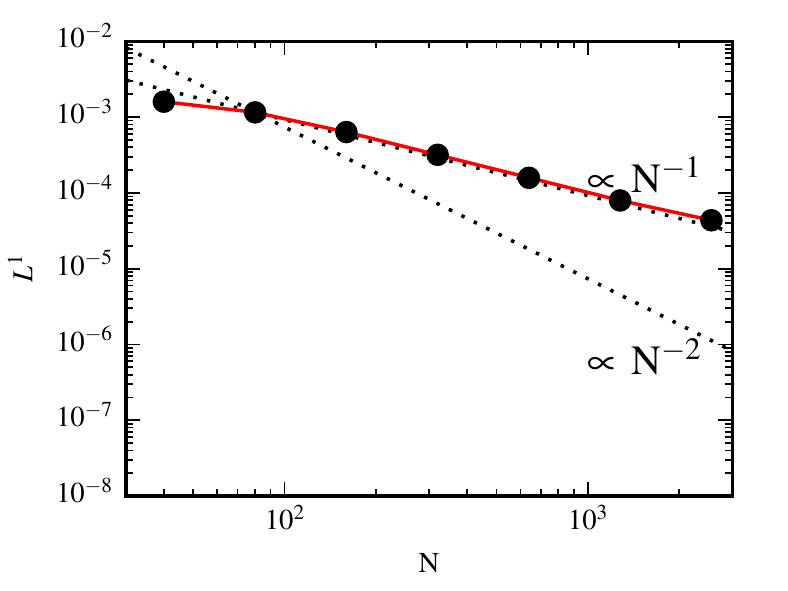}
  \caption{$L^1$ norm of the density field as a function of the
    number of resolution elements per dimension at $t=10$ for the
    isentropic Yee vortex, simulated with the standard \textsc{Arepo}
    implementation using MUSCL-Hancock time integration and
    the improved Green-Gauss gradient estimates.}
  \label{fig:YeeL1}
\end{figure}

In the original \textsc{AREPO} implementation of \citet{Arepo}, all
computations that require geometrical quantities, e.g.~volumes, areas,
or positions of face centroids, use the geometry of the mesh at the
beginning of the timestep. Nevertheless, it has been argued that the
scheme is second-order accurate, based on simple sound-wave tests
\citep{Arepo} using the $L^1$ norm, and based on the much more-demanding
stationary isentropic vortex flow \citep{Yee2000a}, where the error
decreases quadratically with increasing spatial resolution
\citep{Springel2011a}.

In this paper, we define the $L^p$ norm of a continuous field as
\begin{equation}
  L^p = \left( \frac{1}{V} \int_V \left| f \left(\textbf{r} \right) \right|^p  \mathrm{d}V  \right)^{1/p},
\end{equation}
where $f \left(\textbf{r} \right)$ is, for example, the deviation of the density
field from its analytical value at a position $\textbf{r}$.
For our finite volume discretisation this becomes
\begin{equation}
  L^p = \left( \frac{1}{V} \sum_{i=1}^{N_\mathrm{cells}} \left| f_i \right|^p  V_i  \right)^{1/p}.
\end{equation}

The $L^1$ norm of the density field for the Yee vortex (for the detailed
setup, see Appendix~\ref{app:yee}) evolved until $t=10$ with the original
\textsc{AREPO} code is shown in Figure~\ref{fig:YeeL1}. In contrast to 
previous claims, the error only decreases linearly with resolution, i.e.
\textsc{AREPO} shows first order convergence only. The difference to
previous results \citep[in particular][]{Springel2011a} arises from an 
inappropriate definition of the $L^2$-norm employed there.

On static Cartesian meshes, we find that \textsc{Arepo} is fully
second order convergent for smooth problems. On moving meshes,
however, an inaccuracy is introduced since the time integration only
uses the geometry of the mesh at the beginning of the timestep,
ignoring changes of the mesh geometry during a timestep. One
possibility to account for these changes in the time integration is to
do an additional mesh construction at the middle of the timestep and
use the geometrical properties at $t=t_0 + \Delta t / 2$ to calculate
the fluxes. However, this requires at least one further (expensive)
mesh construction per timestep, nearly doubling the computational
cost. Such an approach combined with a Runge-Kutta time integration
scheme is followed in \textsc{Tess} \citep{Duffell2011a}.

\section{Runge-Kutta time integration}

\label{sec:rk}
Alternatively, one can abandon the MUSCL-Hancock scheme and try to use
a different Runge-Kutta time integration scheme that avoids a mid-step
mesh construction. Of particular interest for us is Heun's method,
which is a second-order Runge-Kutta variant that calculates the flux
as an average of the fluxes at the beginning and the end of the
timestep. Applying Heun's method to our moving mesh leads to the
following update equations of the conservative variables for a cell
$i$
\begin{eqnarray}
   \textbf{Q}^\prime_i &=& \textbf{Q}^n_i - \Delta t \sum_j A^n_{ij} \mathbf{\hat{F}}_{ij}^{n} \left( \textbf{U}^n \right)
   \label{eqn:heunorig}, \\
   \textbf{r}^\prime &=& \textbf{r}^n +  \Delta t \, \textbf{w}^n ,\\
   \textbf{Q}^{n+1}_i &=& \textbf{Q}^n_i - \nonumber \\
   &&\frac{\Delta t}{2} \, \left( \sum_j A^n_{ij} \mathbf{\hat{F}}_{ij}^{n} \left( \textbf{U}^n \right) + 
   \sum_j A^\prime_{ij} \mathbf{\hat{F}}_{ij}^{\prime} \left( \textbf{U}^\prime \right) \right) \label{eqn:heunupdate}, \\
   \textbf{r}^{n+1} &=& \textbf{r}^n + \frac{\Delta t}{2} \, \left( \textbf{w}^n + \textbf{w}^\prime \right) .
\end{eqnarray}
Here, the vectors $\textbf{r}$ denote again the coordinates of the
mesh-generating points, $\textbf{w}$ their velocities, and
$\mathbf{\hat{F}}_{ij}$ is an approximation for the fluxes of the
conserved variables over the interface between cells $i$ and $j$. In
principle it is possible to allow the velocities of the
mesh-generating points to change during a timestep
\citep{Duffell2011a}, but we choose to always keep them constant over
the course of a full timestep. 

Note that this update rule also requires the geometry of two different
meshes ($\textbf{r}^n$ and $\textbf{r}^\prime$), and the fluxes have
to be computed twice for every timestep. However, since for Heun's
scheme and a constant velocity $\textbf{w}^\prime = \textbf{w}^n$ of the mesh-generating
points over the whole timestep we can show that
\begin{equation}
\textbf{r}^{n+1} = \textbf{r}^n + \frac{\Delta t}{2} \, \left( \textbf{w}^n + \textbf{w}^\prime \right) = \textbf{r}^n +  \Delta t \, \textbf{w}^n = \textbf{r}^\prime,
\end{equation} 
we can reuse the mesh we constructed for the second half of the current timestep for
the first half of the next timestep. Thus, we only need to construct
the mesh effectively once per timestep, which keeps the mesh
construction effort equal to \textsc{Arepo}'s original
implementation. The fluxes need to be calculated twice as often, but
the computational cost required for this is a relatively small
effort compared to the 3D mesh construction needed in a moving-mesh
code, and therefore does not increase the overall computational cost
significantly.
 
In practice, computing the intermediate state $\textbf{Q}^\prime$
through Eqn.~(\ref{eqn:heunorig}) is inconvenient for several
reasons. In particular, it significantly complicates the internal
bookkeeping of the conserved variables since we need to know
$\textbf{Q}^n_i$ and $\textbf{Q}^\prime_i$ at the same time in
Eqn.~(\ref{eqn:heunupdate}). While this can be resolved in principle,
further complications arise when one tries to consistently implement
this update scheme for individual time steps that vary locally. 

These problems can be circumvented in a simple way by extrapolating 
to the intermediate step using spatial derivatives, like in the previous 
MUSCL-Hancock scheme. This leads to an update of the
conservative variables through 
\begin{eqnarray}
   \textbf{W}^\prime_i &=& \textbf{W}^n_i + \Delta t \, \frac{\partial \textbf{W}}{\partial \textbf{t}}
   \label{eqn:heunnew} , \\
   \textbf{r}^\prime &=& \textbf{r}^n +  \Delta t \, \textbf{w}^n , \\
   \textbf{Q}^{n+1}_i &=& \textbf{Q}^n_i - \\
   &&\frac{\Delta t}{2} \, \left( \sum_j A^n_{ij} \mathbf{\hat{F}}_{ij}^{n} \left( \textbf{W}^n \right) + 
   \sum_j A^\prime_{ij} \mathbf{\hat{F}}_{ij}^{\prime} \left(
      \textbf{W}^\prime \right) \right) ,  \nonumber \\
   \textbf{r}^{n+1} &=& \textbf{r}^\prime.
\end{eqnarray}
Here, we only need to do the time extrapolation for the primitive
variables, and we again use the Euler equations to replace time
derivatives with spatial derivatives. Note that this update scheme can
be easily generalized to local individual time steps. For the
calculation of the fluxes between two cells on different time steps,
the time extrapolation (Eqn.~\ref{eqn:heunnew}) is in this case done
for each cell always from the last time the cell was active. At this time
the estimates for its primitive variables and gradients were obtained. 
Similarly, the linear spatial extrapolation to the current
center of the interface is always done from the center of mass for
which primitive variables and their gradients have been calculated.
Note, that with the time-extrapolation to the intermediate step
our scheme is not a Runge-Kutta scheme anymore, but becomes a
hybrid between Runge-Kutta and MUSCL-Hancock schemes.
 
\section{Least square gradient estimates}
\label{sec:lsf}

Another source of accuracy degradation in the original \textsc{Arepo}
implementation lies in the quality of the gradient estimates, which can
suffer in general problems. The issue appears when a cell is
distorted and its mesh-generating point deviates significantly from
the position of the center of mass of the cell. In this case, the
Voronoi-optimized Green-Gauss estimate of Eqn.~(\ref{eqngradient})
introduces a systematic error, because it assumes that the primitive
variables are known at the mesh-generating points, whereas, by
construction, the primitive variables represent volume-averaged
quantities that represent the value of the underlying
field at the cell's center-of-mass. Although the small steering
motions added by \textsc{Arepo} to the velocities of the mesh
generating points usually manage to keep the mesh nicely regular with
only a small offset between the mesh-generating point and the
center-of-mass of a cell, the typical distance between these two
points can amount to a few percent of the radius of the cell, enough
to significantly degrade the accuracy of the gradient estimate in some
situations.

\begin{figure}
  \centering
  \includegraphics[width=\linewidth]{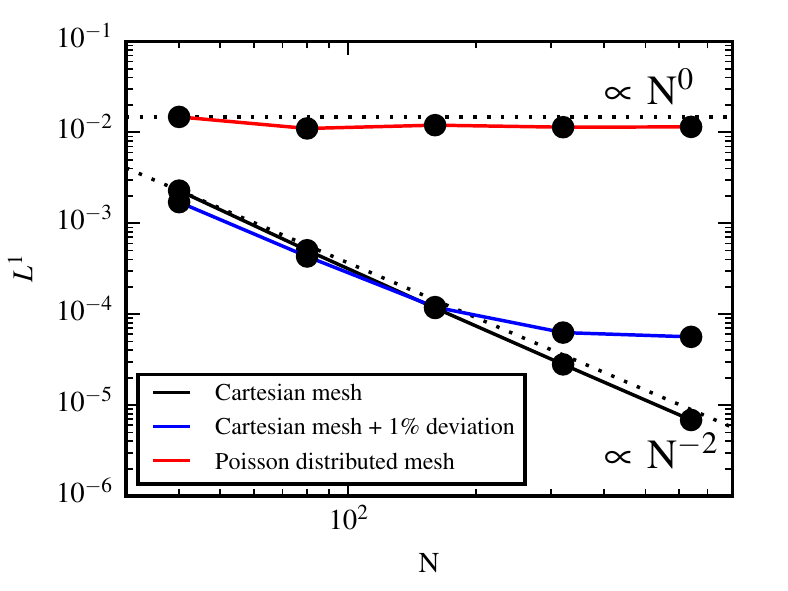}
  \caption{$L^1$ error norm of the Voronoi-optimized Green-Gauss gradient
    estimate of the density field for different types of meshes for
    the initial state of the Yee vortex at $t=0$.}
  \label{fig:YeeGradGG}
\end{figure}
 
We investigate this in Fig.~\ref{fig:YeeGradGG} explicitly, where we
show that such a small deviation already compromises the accuracy of
the gradient estimate, spoiling its convergence rate for higher
resolutions. In fact, a one percent deviation of the
mesh-generating points from the centers of mass in an almost Cartesian
mesh already stops the improvement of the gradient estimate with resolution
for resolutions better than $320^2$ cells. Even worse, for a
completely random mesh in which the mesh-generating points are a
Poisson sample of the simulation domain the gradient estimate does not
improve with resolution at all. Note, however, that the hydro scheme
still converges with first order, even though the error in the gradient estimate
is constant.
 
To more robustly reach second-order convergence of the hydrodynamical
scheme, we therefore need better gradient estimates that are correct
to at least first order for severely distorted meshes where
center-of-mass values are known, equivalent to requiring that linear
gradients are still reproduced exactly in this case. 

\begin{figure}
  \centering
  \includegraphics[width=\linewidth]{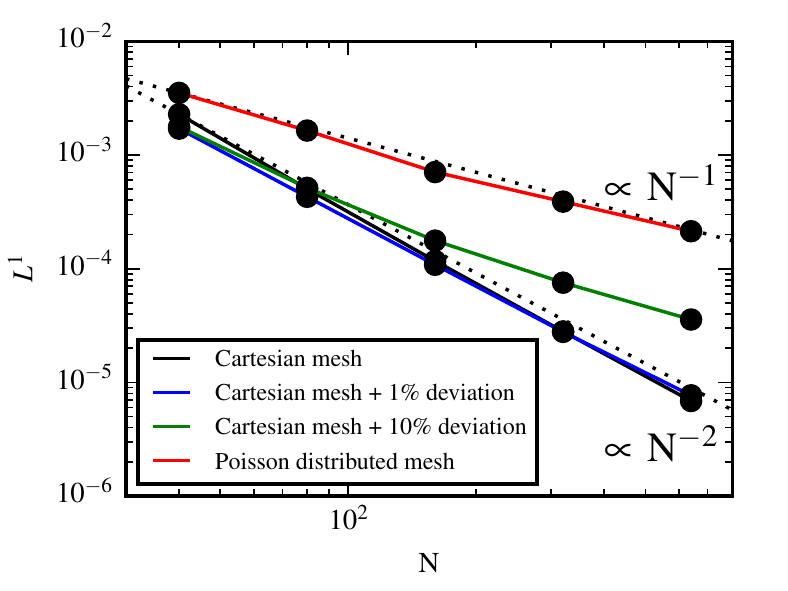}
  \caption{$L^1$ norm of the least-square-fit gradient estimate of 
    the density field for different types of meshes for
    the initial state of the Yee vortex at $t=0$.}
  \label{fig:YeeGradLSF}
\end{figure}

Assuming that we know the local geometry of the mesh and the values of
a quantity $\phi$ at the centroids of the cells, we can obtain such a
gradient estimate through a local least squares fit. We note that
polynomial least squares reconstructions are well known for
unstructured meshes, and form a standard technique in particular for
high-order ENO and WENO schemes
\citep[e.g.][]{OllivierGooch1997}. Also, such methods are well known
for the estimation of gradients \citep[e.g.][]{maron2003a}. They have also been recently used for
SPH and in new variants of mesh-less methods
\citep{HopkinsGizmo} and to compute the cell-centered magnetic
fields in constrained transport schemes for unstructured meshes
\citep{Mocz2014}.  The method assumes that the quantity $\phi$ can be
approximated everywhere within cell $i$ as
\begin{equation}
   \phi \left( \textbf{r} \right) = \phi \left( \textbf{s}_i \right) + \left< \nabla \phi \right>_i \left( \textbf{r} - \textbf{s}_i \right).
\end{equation}
The value of $\phi$ at the center of mass of the cell
$\phi \left( \textbf{s}_i \right) \equiv \phi_i$ 
is known as a prerequisite, 
and we are now looking for
an estimate of the gradient $\left< \nabla \phi \right>_i$. We also
know $\phi$ at the centres of mass of all neighbouring cells $j$, so
we can require that our gradient estimate reproduces those $\phi_j$ as
well as possible. Thus, for every neighbouring cell, we would like to have
\begin{equation}
   \phi_j = \phi_i + \left< \nabla \phi \right>_i \left( \textbf{s}_j - \textbf{s}_i \right).
\end{equation}
For $N$ neighbouring cells this is an overdetermined set of $N$
equations with only up to three free variables (or two in 2D). We
select the best linear approximation for $\left< \nabla \phi \right>_i$ by 
requiring that it minimises the sum of the deviations for all neighbours,
\begin{equation}
   S_\mathrm{{tot}} = \sum_j g_j \left( \phi_j - \phi_i  - \left< \nabla \phi \right>_i \left( \textbf{s}_j - \textbf{s}_i  \right) \right)^2.
\end{equation}
Here, $g_j$ is the relative weight for neighbor $j$. Different choices for these 
weights are possible, but according to our experiments the results are not very 
sensitive to the particular choice made. We follow \citet{Vasconcellos2014} and
set it to $g_j = A_{ij} / \left| \textbf{s}_j - \textbf{s}_i \right|^2$.

To minimise $S_\mathrm{{tot}}$, we use the normal equations which
yield
\begin{eqnarray}
 \hspace*{1cm}  \sum_j g_j \textbf{n}_{ji} \otimes \textbf{n}_{ji} \left< \nabla
  \phi \right>_i \left| \textbf{s}_j - \textbf{s}_i \right|^2  = \nonumber \\
\;\;\;   \sum_j g_j \left( \phi_j - \phi_i \right) \textbf{n}_{ji} \left| \textbf{s}_j - \textbf{s}_i \right|,
\end{eqnarray}
where $\textbf{n}_{ji} = \left( \textbf{s}_j - \textbf{s}_i \right) /
\left| \textbf{s}_j - \textbf{s}_i \right|$. This corresponds to a
$2\times 2$ or $3\times 3$ matrix inversion problem in two- or
three-dimensional problems, respectively, which can be solved by an
appropriate solver to obtain $\left< \nabla \phi \right>_i$.

\begin{figure}
  \centering
  \includegraphics[width=\linewidth]{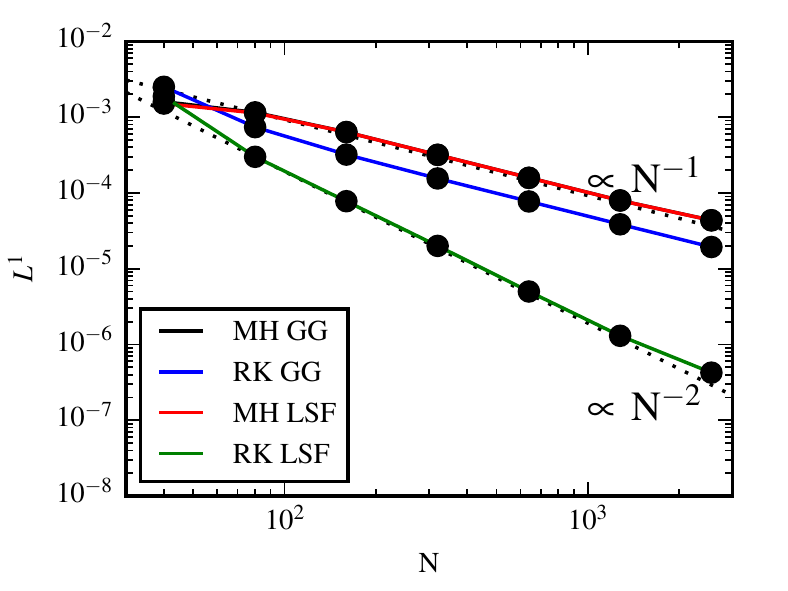}
  \caption{$L^1$ norm of the density field for the evolved Yee vortex at
    $t=10$ when different schemes for the hydrodynamics are employed.
    For the detailed setup of the Yee vortex problem see Appendix~\ref{app:yee}.}
  \label{fig:YeeLRho}
\end{figure}

The accuracy of the new least squares fit gradient estimate is shown in
Fig.~\ref{fig:YeeGradLSF}. On the nearly Cartesian mesh examined
earlier, the gradient estimate is as accurate as the Voronoi-optimized
Green-Gauss estimate. For increasing deviations from the Cartesian
mesh, the accuracy of the least squares gradient estimate slowly
deteriorates, but importantly, it never becomes worse than first
order. In particular, even for the random mesh it still accurately
recovers the linear gradient, unlike the Voronoi-optimized Green-Gauss
estimate.
 
\section{Test problems}
\label{sec:tests}

To examine the difference between the original \textsc{Arepo}
implementation that employs a MUSCL-Hancock time integration
scheme (MH) and the Voronoi-adjusted Green-Gauss gradient estimate (GG)
versus the Runge-Kutta time integrator (RK) combined with the
least squares gradient estimator (LSF), we compare these schemes for several
representative test problems. In obtaining the results for the
different implementations, we always use identical code configurations,
parameters, and initial conditions, and only vary the time integration
method and/or gradient estimate.

All test problems in this paper are run using a Courant factor of $0.3$ to
determine the timestep and employ the same strategy and parameters
to evolve the mesh while keeping it regular. To regularise the mesh, the mesh-generating
points are moved towards the center of mass of their cell. Their total velocity is calculated
as
\begin{equation}
\textbf{v}_{\mathrm{vertex}} = \textbf{v} -\frac{1}{2} \Delta t \frac{\nabla P}{\rho} + \textbf{v}_{\mathrm{reg}}
\end{equation}
where $\textbf{v}$, $P$, $\rho$, $\Delta t$, and $\textbf{v}_{\mathrm{reg}}$ are the fluid velocity, 
pressure, and density, timestep of the cell, and a regularisation component that is added purely to improve
the mesh. It is given by
\begin{equation}
  \textbf{v}_{\mathrm{reg}} = \epsilon \max \left( c, r\, \left| \nabla \times \textbf{v} \right| \right) \frac{ \textbf{d} }{ \left| \textbf{d} \right| }.
\end{equation}
Here, $c$ is the sound speed in the cell, $r$ the approximate extent of the cell 
calculated from its area (volume) in 2D (3D) assuming the cell is a circle (sphere),
and $\textbf{d} = \textbf{r} - \textbf{s}$ is the separation between mesh-generating point 
and center of mass of the cell. Note that this separation is interesting in its own right,
as it can be used to quantify the regularity of the mesh. Moreover, $\epsilon$ is defined as
\begin{equation}
\epsilon = \max \left( 0, \min \left( 0.5, 0.5\, \frac{\alpha - 0.75 \beta}{0.25 \beta} \right) \right)
\end{equation}
and $\alpha$ is given by
\begin{equation}
\alpha = \max_i \left( \frac{1}{N_{\mathrm{dim}}} \frac{2 A_i}{\left| \textbf{r} - \textbf{r}_i \right|} \right),
\end{equation}
where the maximum runs over all neighbours of a cell and their interfaces connecting a cell to them.
We choose the parameter $\beta$ to be 2.25 in all simulations in this paper.

\begin{figure}
  \centering
  \includegraphics[width=\linewidth]{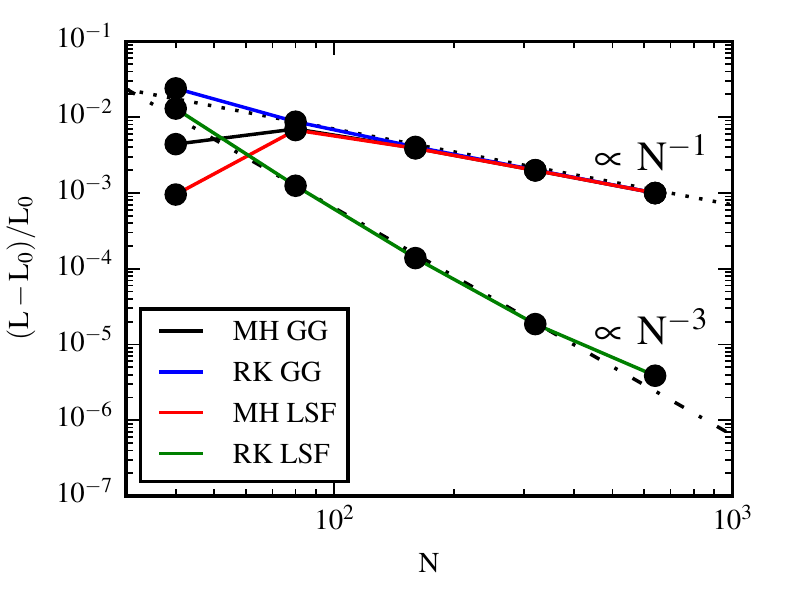}
  \caption{Conservation of total angular momentum for the Yee vortex
    at $t=10$ when different hydrodynamical schemes are used.}
  \label{fig:YeeAngmom}
\end{figure}

\begin{figure}
  \centering
  \includegraphics[width=\linewidth]{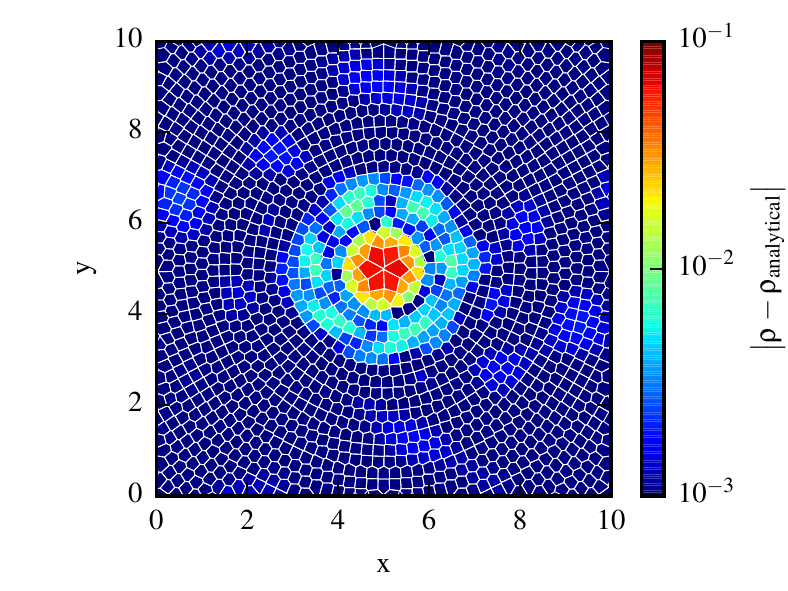}
  \includegraphics[width=\linewidth]{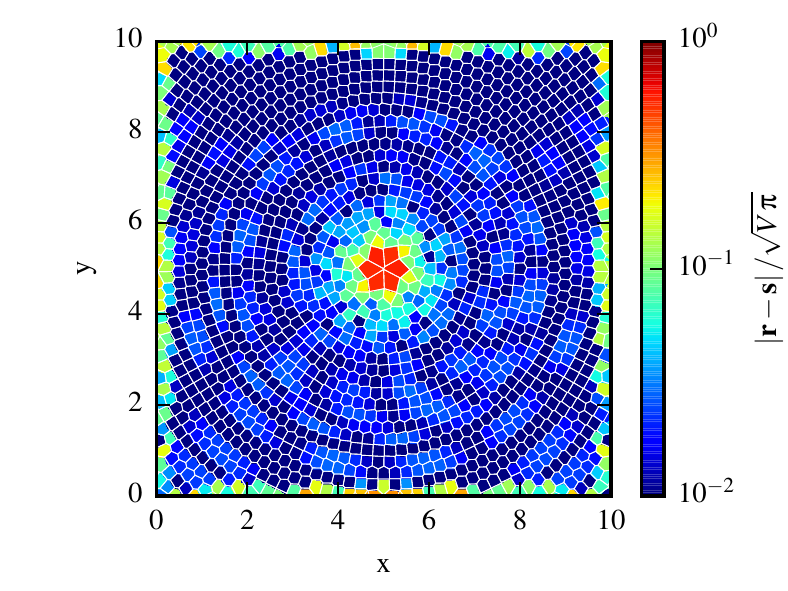}
  \caption{Density error compared to the analytical solution (top panel) and relative separation 
  between mesh-generating point and center of mass (bottom panel) at $t=100$ for the 
  RKLSF Yee vortex.}
  \label{fig:YeeMesh}
\end{figure}

\subsection{Yee vortex}
 
We first reevaluate the convergence rate for the isentropic Yee
vortex. Since this vortex flow is completely smooth, we in principle
expect to see second-order convergence in the strong $L^1$ norm if
the implementation is correct. As shown in Fig.~\ref{fig:YeeLRho},
we indeed obtain second-order convergence up to very high resolution
for \textsc{Arepo} in the moving-mesh case if we use {\em both} the
least squares gradient estimate and the Runge-Kutta time integration
scheme. Using only one of the two improvements while keeping the
MUSCL-Hancock integration scheme or the improved Green-Gauss gradient
estimate, respectively, does not change the convergence rate at all
and makes the implementation drop back to first-order convergence in
the $L^1$ norm. The detailed setup including the initial
mesh can be found in Appendix~\ref{app:yee}.

\begin{figure*}
  \centering
  \includegraphics[width=0.99\linewidth]{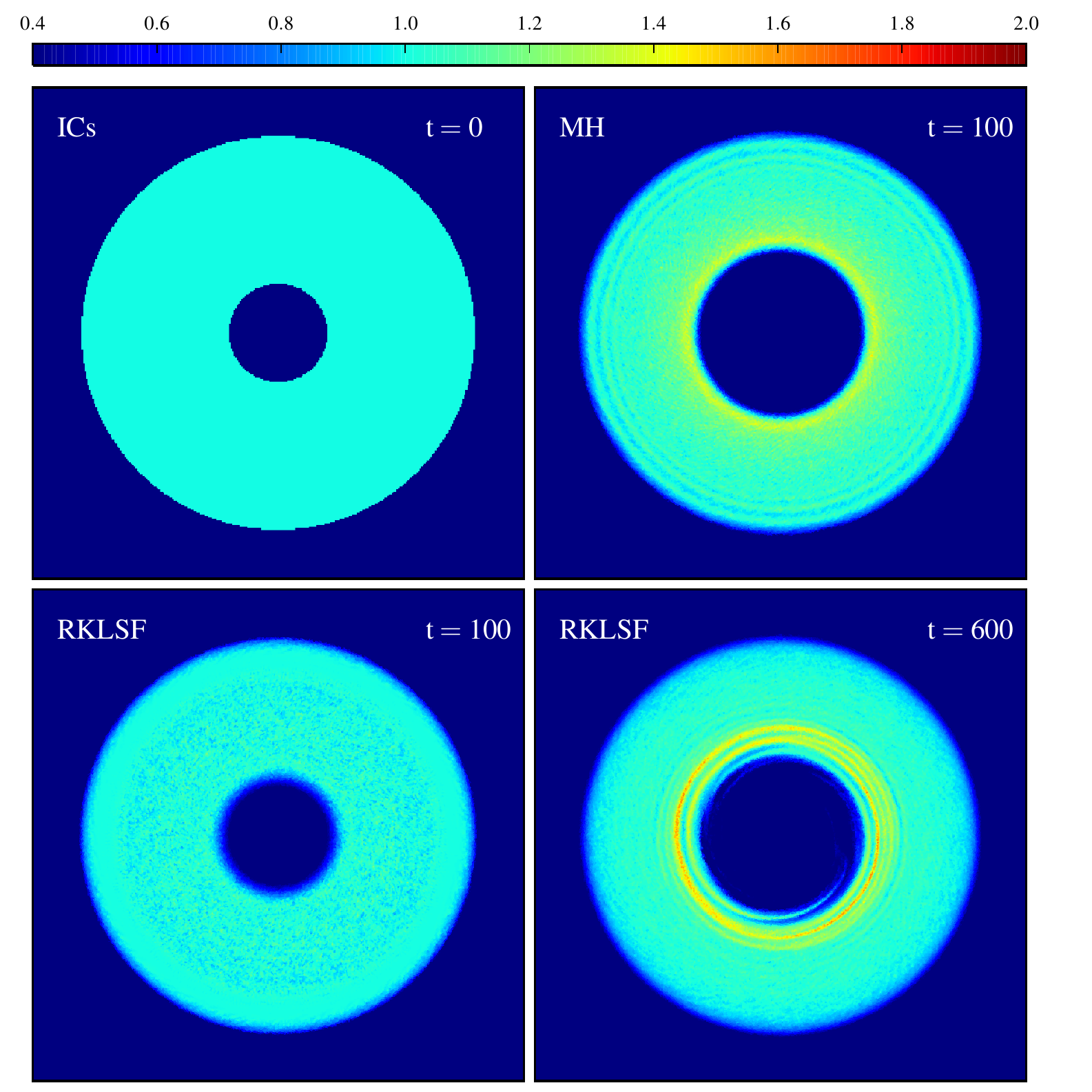}
  \caption{Evolution of the surface density of a two-dimensional cold
    Keplerian disc. The top panels show the initial conditions and the
    evolved state at $t=100$ obtained with the original \textsc{Arepo}
    implementation based on the improved Green-Gauss gradient
    estimate and the MUSCL-Hancock time integration. The bottom panels
    show the new implementation with least squares gradient estimate
    and Runge-Kutta time integration at $t=100$ and $t=600$,
    respectively. At the latter time, the inner disc has finished more
    than $250$ orbits. The detailed setup can be found in 
    Appendix~\ref{app:disk}.}
  \label{fig:KeplerianDisk}
\end{figure*}
 
It is interesting to note that the \textsc{Tess} code has been shown
to be second-order accurate for an isentropic sound wave with a
Runge-Kutta time integration scheme \citep{Duffell2011a}, while the same
Voronoi-optimized Green-Gauss gradient estimate as implemented
originally in \textsc{Arepo} has been used. With this gradient
estimate, the convergence order is however expected to break down for 
meshes in which the mesh-generating points are not close to the centers 
of mass of their cells. The difference here is
most likely that the one-dimensional propagation of the wave in this
particular problem only stretches the mesh, without displacing the
mesh-generating points from the centres of mass of their cells. As
discussed before, in this special case the Voronoi-optimized gradient
estimate is sufficient and does not negatively impact the convergence
properties. For the more demanding Yee vortex this is not the case,
however, and the convergence is expected to break down in
\textsc{Tess} just as in the original \textsc{Arepo} code.
 
Another interesting observation to make is that second-order
convergence in \textsc{Arepo} for the new implementation as measured
by computing the $L^1$ norm between the current density field and the
analytical solution is only reached at very high resolution when the
initial mesh-generating points are set up in rings around the center
of the vortex. If an initially Cartesian mesh is adopted instead, the 
convergence degrades to first order at sufficiently high resolution.
As shown in Fig.~\ref{fig:YeeMesh} there is only a weak correlation
between the density error in a cell and its distortion, as both are largest
close to the center of the vortex.

This is fundamentally caused by the discretization of the analytical
problem onto our mesh. There are two discretizations involved, a first
discreziation on the initial mesh to generate the discretized initial
conditions and a second discretization on the current mesh when we
measure the $L^1$ norm. If the structure of the mesh changes
systematically between the initial and the current mesh, there is also
a systematic difference in the two discretizations, which turns out to
lead to a first order error that dominates the total error at
sufficiently high resolution.  Because an initially Cartesian mesh
will eventually be transformed to a spherical mesh by the mesh
regularization, the measured $L^1$ norm after the mesh changed will
not be better than first order at high resolution.  To overcome this
problem, we arrange the mesh-generating points in the initial mesh
already on circles around the center of the vortex. This mesh
configuration is stable for the problem, thus there is no systematic
change in the discretisation over time.

Another approach is to look at error measures intrinsic to the
discrete problem that do not require a second discretisation of the
analytical problem.  Analysing the conservation of total angular
momentum (which also may be viewed as a norm) instead of the $L^1$
norm of the total density field is therefore considerably simpler and
also more robust. In particular, here the measured convergence rate is
independent of the initial mesh, and the rearrangement from an
initially Cartesian to a circular mesh does not give rise to
additional errors. As shown in Fig.~\ref{fig:YeeAngmom}, the angular
momentum conservation is again only improved if {\em both}, a better
time integration method and a more accurate gradient estimate, are
used. Interestingly, the angular momentum even shows third order
convergence in the new implementation, although this may be the result
of the special spherical symmetry of the vortex flow and may be lost
for more general problems.

\subsection{Keplerian disc}

Evolving a cold Keplerian disc for many orbits is a common problem in
astrophysics, with applications from planetary to galactic
discs. Using such a set-up in the limit of a pressure-less gas disc is
a demanding test problem \citep{Cullen2010, HopkinsGizmo}. In fact, as
highlighted by \citet{HopkinsGizmo}, many state-of-the-art static mesh
or SPH codes have severe problems in coping accurately with this
situation. Instead, the disc is typically destroyed during the first
$10$ orbits by these methods.

The original \textsc{AREPO} code with the MUSCL-Hancock time
integration and improved Green-Gauss gradients already does quite well
on this problem, as shown in Fig.~\ref{fig:KeplerianDisk} where we
display the disc after roughly $15$ inner orbits with a resolution of
320x320 cells. However, whereas the
outer parts of the disc are very stable, the inner boundary of the
disc moves outwards with time owing to systematic errors in angular
momentum conservation. Our new implementation with Runge-Kutta time
integration and least squares gradients works significantly better,
keeping the disc essentially perfectly stable until $t=100$, and
showing similar errors only much later, at times $t \simeq 600$ or
later. These residual errors can now be much more efficiently improved
to essentially arbitrary precision by an increase of the spatial
resolution, thanks to the improved convergence order. This is in line
with previous results on conservation of angular momentum in
\textsc{AREPO} for more realistic problems \citep{Munoz2015}.

These results compare very favourably not only to stationary 
Eulerian grid codes on Cartesian meshes, but also to the new 
mesh-free hydrodynamical method proposed by \citet{HopkinsGizmo}.
Note, however, that the optimal mesh configuration for this specific
problem is most likely a polar grid.

\section{Testing on realistic applications}  
\label{sec:world}

\subsection{Evolution of a stellar binary}

In light of our previous results, it is interesting to investigate
whether the improvements in the \textsc{Arepo} code proposed here
affect the results of real world scientific applications of the code.
Compared to the Yee vortex and the Keplerian disc, which are smooth
hydrodynamics-only problems, or hydrodynamical problems with a
constant external gravitational field, one example of a more complex
problem including discontinuities, shocks, and self-gravity is the
inspiral and merger of a close binary system of two white dwarfs.

\begin{figure}
  \centering
  \includegraphics[width=\linewidth]{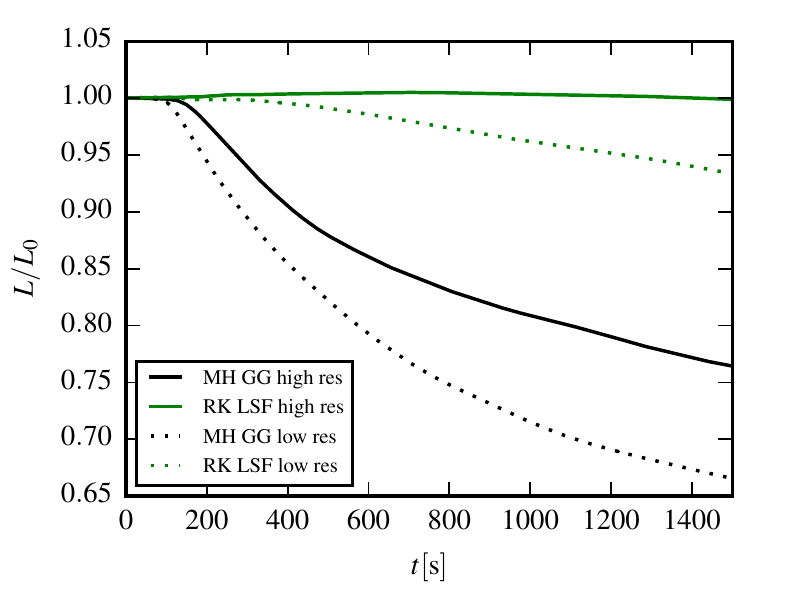}
  \caption{Conservation of total angular momentum for the merger of
    two white dwarfs with an initial orbital period of $44$s for the
    MUSCL-Hancock time integration and improved Green-Gauss gradient
    estimate and the Runge-Kutta time integration combined with the
    least squares gradient estimate.}
  \label{fig:wdmerger}
\end{figure}

Such simulations are essential to understanding the complex evolution
of the merger process and to determining the fate of the binary
system \citep[see, e.g.][]{Zhu2013,Pakmor2013SN,Dan2014}.
The initial mesh for the white dwarfs is constructed from shells that are
tesselated using the HEALPIX algorithm to obtain regular cells with
roughly the same mass \citep{Pakmor2012SG}. The simulation box is 
then filled by a low resolution Cartesian background grid.
Figure~\ref{fig:wdmerger} shows the total angular momentum for the 
merger of two white dwarfs with masses of $0.65\mathrm{M_\odot}$ and 
$0.625\mathrm{M_\odot}$, an initial orbital period of $44$\,s, identical to 
the setup of the same system in \citet{Zhu2013}. We employ a mass resolution 
of $10^{-6}\mathrm{M_\odot}$ (high resolution) and $5\,10^{-6}\mathrm{M_\odot}$
(low resolution), respectively.

The binary merges at around $t=200\,{\rm s}$, after initial mass
transfer reduced the separation and lead to runaway mass transfer. It then forms a
differentially rotating merger remnant. The differential rotation in
the merger remnant is crucial for its further evolution. As shown in
Figure~\ref{fig:wdmerger}, conservation of angular momentum during the
further evolution of the merger remnant is significantly violated for
the implementation with MUSCL-Hancock time integration and improved
Green-Gauss gradient estimates. The system loses a significant
fraction of the total angular momentum present in the simulation. In
contrast, the new implementation with Runge-Kutta time integration and
least squares gradient estimates conserves the total angular momentum
in the simulation to a relative error of about one percent even after
many orbits for the high resolution simulation and still loses less than
$10\%$ of the total angular momentum in the low resolution simulation.

\begin{figure*}
  \centering
  \includegraphics[width=0.99\linewidth]{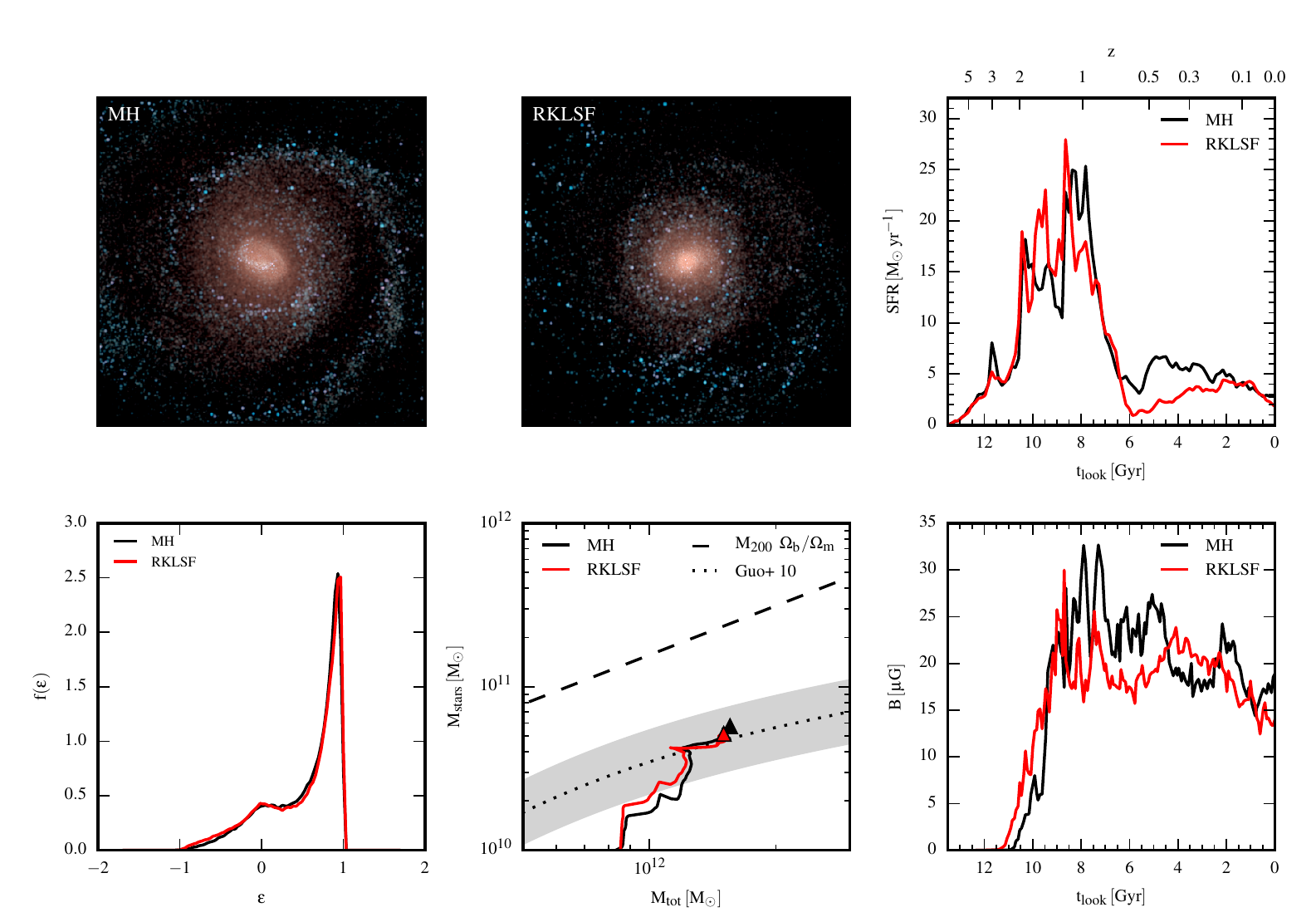}
  \caption{Comparison of two cosmological zoom runs of a galaxy
    similar to the Milky-Way \citep[as in][]{Marinacci2013} using
    either the combination of MUSCL-Hancock time integration and
    improved Green-Gauss gradient estimate, or the combination of
    Runge-Kutta time integration and least squares gradient estimate,
    respectively. Initial conditions, code configuration and
    parameters including all sub-grid physics are identical for both
    runs and chosen in line with \citet{Marinacci2013} and
    \citet{Pakmor2014}.  The top row shows, from left to right,
    stellar projections of the two runs at $z=0$ and the evolution of
    the star formation rate in the disc of the main galaxy. The bottom
    row shows circularities of stars in the disc, evolution of the
    total mass of the halo compared to its stellar mass, and the
    average root mean square magnetic field strength in the disc for
    the two runs.}
  \label{fig:galaxy}
\end{figure*}

\subsection{Cosmological zoom simulation of galaxy formation}
 
An even more complex problem are simulations of galaxy formation
and evolution. They not only involve
self-gravity, but also feature high Mach numbers and turbulent flows, as
well as a large number of additional source terms to model effects
like radiative cooling of gas or the energy injection of evolving
stars. In addition, sometimes explicit sub-grid models for unresolved
physics are used that are introduced as a modified equation of state or
pressure floors and the like.

To compare the performance of our new code with the original
\textsc{Arepo} version in this regime we have repeated one of the
recent Milky Way galaxy formation simulations of \citet{Marinacci2013}
and \citet{Pakmor2014}. These are advanced calculations that can serve
as an example for current state of the art simulations of cosmic
structure formation
\citep[see, e.g.][]{Agertz2011,Scannapieco2012,Stinson2013,Hopkins2014FIRE,Vogelsberger2014Illustris,Schaye2015Eagle,Khandai2015}.
As shown in the results overview of
Fig.~\ref{fig:galaxy}, there is no significant difference between the
results obtained for both implementations.
 
A possible interpretation of the lack of differences for the
cosmological runs is that the properties of galaxies and their
dynamics are already captured sufficiently accurately by the standard
\textsc{Arepo} implementation. The accuracy improvements brought about
by the new formulation proposed here are either irrelevant for this
problem, or are completely dominated by first order errors introduced
by the sub-grid models for radiative cooling, star formation, and
feedback. The latter is particularly likely as it is now well
understood that the outcome of galaxy formation simulations depends
very sensitively on the treatment of highly non-linear feedback
processes. Thus, in regions of the simulation that are crucially
shaped by feedback we do not expect our improvements to the code to
lead to significant changes of the results. Changing this to improve the
solution requires
implementing and coupling the source terms such that the combined system
is second order accurate.
This may be different in regions where hydrodynamics and gravity are the
only relevant phenomena. For example, conceivably some differences may 
be found in the approximately hydrostatic atmosphere of rich galaxy clusters,
where our new formulation may result in a marginally better
representation of turbulence.

\section{Conclusions}
\label{sec:conclusion}

In this paper, we have discussed in detail two simple modifications of
the cosmological moving-mesh code \textsc{Arepo}, which are
nevertheless quite important to recover full second-order convergence
in the $L^1$ norm for general smooth problems with non-trivial
mesh motions. One of these changes concerns the time integration, where
a Runge-Kutta time integration scheme is adopted instead of the MUSCL-Hancock
approach in order to account for changes of the mesh geometry during a 
timestep at second order. This does not increase the number of mesh-constructions
needed, but does double the number of required flux computations. The other
change is the adoption of a more general (and slightly more expensive)
gradient estimate that retains the necessary accuracy even for large
offsets between the mesh-generating points and the centers of masses
of cells.

These improvements are most relevant for smooth, pure hydrodynamics
problems where the influence of self-gravity and of complicated
source terms is limited.  Among the astrophysical problems that fall
into this regime and that have already been tackled with the original
version of \textsc{Arepo} are proto-planetary discs \citep{Munoz2014},
cold gas in galactic discs \citep{Smith2014}, and dynamical stellar
mergers \citep{Pakmor2013SN}. Such simulations and similar problems
will benefit in the future from the added accuracy facilitated by the
improvements proposed here. It is also prudent to carry out additional
tests to see whether previous simulations were sometimes degradated in
a noticeable way by an accuracy loss of the code, for example by an
unnecessarily large error in the conservation of angular momentum. We
expect such noticeable errors only for problems that evolve a system
for many dynamical timescales.

Cosmological simulations of galaxy formation seem to be
unaffected by the improvements of the hydrodynamical moving-mesh
scheme proposed here. This is of course reassuring as it means that
previous results from galaxy formation studies carried out with
\textsc{Arepo}, both with zoom-in techniques
\citep{Marinacci2013,Pakmor2014} and in large cosmological boxes
\citep{Vogelsberger2014Illustris,Genel2014}, have not suffered from
the convergence rate issues discussed above. Nevertheless, it is
clearly desirable to use the improved scheme in the future in this
regime as well, especially since it offers higher accuracy at an
insignificant increase of the computational cost.

\section*{Acknowledgements}
    
This work has been supported by the European Research Council under
ERC-StG grant EXAGAL-308037 and by the Klaus Tschira Foundation. VS
and KS acknowledge support through subproject EXAMAG of the Priority
Programme 1648 ``Software for Exascale Computing'' of the German
Science Foundation. Part of the simulations of this paper used the
SuperMUC system at the Leibniz Computing Centre, Garching.
PM acknowledges support by the National Science Foundation Graduate 
Research Fellowship under grant no. DGE-1144152. STO acknowledges
financial support from Studienstiftung des deutschen Volkes. AB and KS
acknowledge support by the IMPRS for Astronomy \& Cosmic Physics at
the University of Heidelberg.
    
\bibliographystyle{mnras}

\begin{thebibliography}{}
\makeatletter
\relax
\def\mn@urlcharsother{\let\do\@makeother \do\$\do\&\do\#\do\^\do\_\do\%\do\~}
\def\mn@doi{\begingroup\mn@urlcharsother \@ifnextchar [ {\mn@doi@}
  {\mn@doi@[]}}
\def\mn@doi@[#1]#2{\def\@tempa{#1}\ifx\@tempa\@empty \href
  {http://dx.doi.org/#2} {doi:#2}\else \href {http://dx.doi.org/#2} {#1}\fi
  \endgroup}
\def\mn@eprint#1#2{\mn@eprint@#1:#2::\@nil}
\def\mn@eprint@arXiv#1{\href {http://arxiv.org/abs/#1} {{\tt arXiv:#1}}}
\def\mn@eprint@dblp#1{\href {http://dblp.uni-trier.de/rec/bibtex/#1.xml}
  {dblp:#1}}
\def\mn@eprint@#1:#2:#3:#4\@nil{\def\@tempa {#1}\def\@tempb {#2}\def\@tempc
  {#3}\ifx \@tempc \@empty \let \@tempc \@tempb \let \@tempb \@tempa \fi \ifx
  \@tempb \@empty \def\@tempb {arXiv}\fi \@ifundefined
  {mn@eprint@\@tempb}{\@tempb:\@tempc}{\expandafter \expandafter \csname
  mn@eprint@\@tempb\endcsname \expandafter{\@tempc}}}

\bibitem[\protect\citeauthoryear{{Agertz}, {Teyssier}  \& {Moore}}{{Agertz}
  et~al.}{2011}]{Agertz2011}
{Agertz} O.,  {Teyssier} R.,   {Moore} B.,  2011, \mn@doi [\mnras]
  {10.1111/j.1365-2966.2010.17530.x}, \href
  {http://adsabs.harvard.edu/abs/2011MNRAS.410.1391A} {410, 1391}

\bibitem[\protect\citeauthoryear{{Almgren} et~al.,}{{Almgren}
  et~al.}{2010}]{Castro}
{Almgren} A.~S.,  et~al., 2010, \mn@doi [\apj] {10.1088/0004-637X/715/2/1221},
  \href {http://adsabs.harvard.edu/abs/2010ApJ...715.1221A} {715, 1221}

\bibitem[\protect\citeauthoryear{{Bryan} et~al.,}{{Bryan} et~al.}{2014}]{ENZO}
{Bryan} G.~L.,  et~al., 2014, \mn@doi [\apjs] {10.1088/0067-0049/211/2/19},
  \href {http://adsabs.harvard.edu/abs/2014ApJS..211...19B} {211, 19}

\bibitem[\protect\citeauthoryear{{Cullen} \& {Dehnen}}{{Cullen} \&
  {Dehnen}}{2010}]{Cullen2010}
{Cullen} L.,  {Dehnen} W.,  2010, \mn@doi [\mnras]
  {10.1111/j.1365-2966.2010.17158.x}, \href
  {http://adsabs.harvard.edu/abs/2010MNRAS.408..669C} {408, 669}

\bibitem[\protect\citeauthoryear{{Dan}, {Rosswog}, {Br{\"u}ggen}  \&
  {Podsiadlowski}}{{Dan} et~al.}{2014}]{Dan2014}
{Dan} M.,  {Rosswog} S.,  {Br{\"u}ggen} M.,   {Podsiadlowski} P.,  2014,
  \mn@doi [\mnras] {10.1093/mnras/stt1766}, \href
  {http://adsabs.harvard.edu/abs/2014MNRAS.438...14D} {438, 14}

\bibitem[\protect\citeauthoryear{{Duffell} \& {MacFadyen}}{{Duffell} \&
  {MacFadyen}}{2011}]{Duffell2011a}
{Duffell} P.~C.,  {MacFadyen} A.~I.,  2011, \mn@doi [\apjs]
  {10.1088/0067-0049/197/2/15}, \href
  {http://adsabs.harvard.edu/abs/2011ApJS..197...15D} {197, 15}

\bibitem[\protect\citeauthoryear{{Duffell} \& {MacFadyen}}{{Duffell} \&
  {MacFadyen}}{2012}]{Duffell2012}
{Duffell} P.~C.,  {MacFadyen} A.~I.,  2012, \mn@doi [\apj]
  {10.1088/0004-637X/755/1/7}, \href
  {http://adsabs.harvard.edu/abs/2012ApJ...755....7D} {755, 7}

\bibitem[\protect\citeauthoryear{{Fryxell} et~al.,}{{Fryxell}
  et~al.}{2000}]{FLASH}
{Fryxell} B.,  et~al., 2000, \mn@doi [\apjs] {10.1086/317361}, \href
  {http://adsabs.harvard.edu/abs/2000ApJS..131..273F} {131, 273}

\bibitem[\protect\citeauthoryear{{Genel} et~al.,}{{Genel}
  et~al.}{2014}]{Genel2014}
{Genel} S.,  et~al., 2014, \mn@doi [\mnras] {10.1093/mnras/stu1654}, \href
  {http://adsabs.harvard.edu/abs/2014MNRAS.445..175G} {445, 175}

\bibitem[\protect\citeauthoryear{{Gnedin}}{{Gnedin}}{1995}]{gnedin1995a}
{Gnedin} N.~Y.,  1995, \mn@doi [\apjs] {10.1086/192141}, \href
  {http://adsabs.harvard.edu/abs/1995ApJS...97..231G} {97, 231}

\bibitem[\protect\citeauthoryear{{Greif}, {Springel}, {White}, {Glover},
  {Clark}, {Smith}, {Klessen}  \& {Bromm}}{{Greif} et~al.}{2011}]{Greif2011}
{Greif} T.~H.,  {Springel} V.,  {White} S.~D.~M.,  {Glover} S.~C.~O.,  {Clark}
  P.~C.,  {Smith} R.~J.,  {Klessen} R.~S.,   {Bromm} V.,  2011, \mn@doi [\apj]
  {10.1088/0004-637X/737/2/75}, \href
  {http://adsabs.harvard.edu/abs/2011ApJ...737...75G} {737, 75}

\bibitem[\protect\citeauthoryear{{Greif}, {Bromm}, {Clark}, {Glover}, {Smith},
  {Klessen}, {Yoshida}  \& {Springel}}{{Greif} et~al.}{2012}]{Greif2012}
{Greif} T.~H.,  {Bromm} V.,  {Clark} P.~C.,  {Glover} S.~C.~O.,  {Smith} R.~J.,
   {Klessen} R.~S.,  {Yoshida} N.,   {Springel} V.,  2012, \mn@doi [\mnras]
  {10.1111/j.1365-2966.2012.21212.x}, \href
  {http://adsabs.harvard.edu/abs/2012MNRAS.424..399G} {424, 399}

\bibitem[\protect\citeauthoryear{{Guillard} \& {Farhat}}{{Guillard} \&
  {Farhat}}{2000}]{guillard2000a}
{Guillard} H.,  {Farhat} C.,  2000, Computer Methods in Applied Mechanics and
  Engineering, 190, 1467

\bibitem[\protect\citeauthoryear{Hirt, {Amsden}  \& {Cook}}{Hirt
  et~al.}{1974}]{hirt1974}
Hirt C.~W.,  {Amsden} A.~A.,   {Cook} J.~L.,  1974, JCP, 14, 227

\bibitem[\protect\citeauthoryear{{Hopkins}}{{Hopkins}}{2014}]{HopkinsGizmo}
{Hopkins} P.~F.,  2014, preprint, \href
  {http://adsabs.harvard.edu/abs/2014arXiv1409.7395H} {} (\mn@eprint {arXiv}
  {1409.7395})

\bibitem[\protect\citeauthoryear{{Hopkins}, {Kere{\v s}}, {O{\~n}orbe},
  {Faucher-Gigu{\`e}re}, {Quataert}, {Murray}  \& {Bullock}}{{Hopkins}
  et~al.}{2014}]{Hopkins2014FIRE}
{Hopkins} P.~F.,  {Kere{\v s}} D.,  {O{\~n}orbe} J.,  {Faucher-Gigu{\`e}re}
  C.-A.,  {Quataert} E.,  {Murray} N.,   {Bullock} J.~S.,  2014, \mn@doi
  [\mnras] {10.1093/mnras/stu1738}, \href
  {http://adsabs.harvard.edu/abs/2014MNRAS.445..581H} {445, 581}

\bibitem[\protect\citeauthoryear{{Khandai}, {Di Matteo}, {Croft}, {Wilkins},
  {Feng}, {Tucker}, {DeGraf}  \& {Liu}}{{Khandai} et~al.}{2015}]{Khandai2015}
{Khandai} N.,  {Di Matteo} T.,  {Croft} R.,  {Wilkins} S.,  {Feng} Y.,
  {Tucker} E.,  {DeGraf} C.,   {Liu} M.-S.,  2015, \mn@doi [\mnras]
  {10.1093/mnras/stv627}, \href
  {http://adsabs.harvard.edu/abs/2015MNRAS.450.1349K} {450, 1349}

\bibitem[\protect\citeauthoryear{{Marinacci}, {Pakmor}  \&
  {Springel}}{{Marinacci} et~al.}{2013}]{Marinacci2013}
{Marinacci} F.,  {Pakmor} R.,   {Springel} V.,  2013, \mn@doi [\mnras]
  {10.1093/mnras/stt2003}, \href
  {http://adsabs.harvard.edu/abs/2013MNRAS.tmp.2660M} {}

\bibitem[\protect\citeauthoryear{{Maron} \& {Howes}}{{Maron} \&
  {Howes}}{2003}]{maron2003a}
{Maron} J.~L.,  {Howes} G.~G.,  2003, \mn@doi [\apj] {10.1086/377296}, \href
  {http://adsabs.harvard.edu/abs/2003ApJ...595..564M} {595, 564}

\bibitem[\protect\citeauthoryear{{Mocz}, {Vogelsberger}  \& {Hernquist}}{{Mocz}
  et~al.}{2014}]{Mocz2014}
{Mocz} P.,  {Vogelsberger} M.,   {Hernquist} L.,  2014, \mn@doi [\mnras]
  {10.1093/mnras/stu865}, \href
  {http://adsabs.harvard.edu/abs/2014MNRAS.442...43M} {442, 43}

\bibitem[\protect\citeauthoryear{{Mu{\~n}oz}, {Kratter}, {Springel}  \&
  {Hernquist}}{{Mu{\~n}oz} et~al.}{2014}]{Munoz2014}
{Mu{\~n}oz} D.~J.,  {Kratter} K.,  {Springel} V.,   {Hernquist} L.,  2014,
  \mn@doi [\mnras] {10.1093/mnras/stu1918}, \href
  {http://adsabs.harvard.edu/abs/2014MNRAS.445.3475M} {445, 3475}

\bibitem[\protect\citeauthoryear{{Mu{\~n}oz}, {Kratter}, {Vogelsberger},
  {Hernquist}  \& {Springel}}{{Mu{\~n}oz} et~al.}{2015}]{Munoz2015}
{Mu{\~n}oz} D.~J.,  {Kratter} K.,  {Vogelsberger} M.,  {Hernquist} L.,
  {Springel} V.,  2015, \mn@doi [\mnras] {10.1093/mnras/stu2220}, \href
  {http://adsabs.harvard.edu/abs/2015MNRAS.446.2010M} {446, 2010}

\bibitem[\protect\citeauthoryear{Ollivier-Gooch}{Ollivier-Gooch}{1997}]{OllivierGooch1997}
Ollivier-Gooch C.~F.,  1997, Journal of Computational Physics, 133, 6

\bibitem[\protect\citeauthoryear{{Pakmor}, {Edelmann}, {R{\"o}pke}  \&
  {Hillebrandt}}{{Pakmor} et~al.}{2012}]{Pakmor2012SG}
{Pakmor} R.,  {Edelmann} P.,  {R{\"o}pke} F.~K.,   {Hillebrandt} W.,  2012,
  \mn@doi [\mnras] {10.1111/j.1365-2966.2012.21383.x}, \href
  {http://adsabs.harvard.edu/abs/2012MNRAS.424.2222P} {424, 2222}

\bibitem[\protect\citeauthoryear{{Pakmor}, {Kromer}, {Taubenberger}  \&
  {Springel}}{{Pakmor} et~al.}{2013}]{Pakmor2013SN}
{Pakmor} R.,  {Kromer} M.,  {Taubenberger} S.,   {Springel} V.,  2013, \mn@doi
  [\apjl] {10.1088/2041-8205/770/1/L8}, \href
  {http://adsabs.harvard.edu/abs/2013ApJ...770L...8P} {770, L8}

\bibitem[\protect\citeauthoryear{{Pakmor}, {Marinacci}  \& {Springel}}{{Pakmor}
  et~al.}{2014}]{Pakmor2014}
{Pakmor} R.,  {Marinacci} F.,   {Springel} V.,  2014, \mn@doi [\apjl]
  {10.1088/2041-8205/783/1/L20}, \href
  {http://adsabs.harvard.edu/abs/2014ApJ...783L..20P} {783, L20}

\bibitem[\protect\citeauthoryear{{Pen}}{{Pen}}{1998}]{pen1998a}
{Pen} U.-L.,  1998, \mn@doi [\apjs] {10.1086/313074}, \href
  {http://adsabs.harvard.edu/abs/1998ApJS..115...19P} {115, 19}

\bibitem[\protect\citeauthoryear{{Scannapieco}, {Wadepuhl}, {Parry}, {Navarro},
  {Jenkins}, {Springel}, {Teyssier}  \& et al.}{{Scannapieco}
  et~al.}{2012}]{Scannapieco2012}
{Scannapieco} C.,  {Wadepuhl} M.,  {Parry} O.~H.,  {Navarro} J.~F.,  {Jenkins}
  A.,  {Springel} V.,  {Teyssier} R.,   et al. 2012, \mn@doi [\mnras]
  {10.1111/j.1365-2966.2012.20993.x}, \href
  {http://esoads.eso.org/abs/2012MNRAS.423.1726S} {423, 1726}

\bibitem[\protect\citeauthoryear{{Schaye} et~al.,}{{Schaye}
  et~al.}{2015}]{Schaye2015Eagle}
{Schaye} J.,  et~al., 2015, \mn@doi [\mnras] {10.1093/mnras/stu2058}, \href
  {http://adsabs.harvard.edu/abs/2015MNRAS.446..521S} {446, 521}

\bibitem[\protect\citeauthoryear{{Sijacki}, {Vogelsberger}, {Kere{\v s}},
  {Springel}  \& {Hernquist}}{{Sijacki} et~al.}{2012}]{Sijacki2012}
{Sijacki} D.,  {Vogelsberger} M.,  {Kere{\v s}} D.,  {Springel} V.,
  {Hernquist} L.,  2012, \mn@doi [\mnras] {10.1111/j.1365-2966.2012.21466.x},
  \href {http://adsabs.harvard.edu/abs/2012MNRAS.424.2999S} {424, 2999}

\bibitem[\protect\citeauthoryear{{Smith}, {Glover}, {Clark}, {Klessen}  \&
  {Springel}}{{Smith} et~al.}{2014}]{Smith2014}
{Smith} R.~J.,  {Glover} S.~C.~O.,  {Clark} P.~C.,  {Klessen} R.~S.,
  {Springel} V.,  2014, \mn@doi [\mnras] {10.1093/mnras/stu616}, \href
  {http://adsabs.harvard.edu/abs/2014MNRAS.441.1628S} {441, 1628}

\bibitem[\protect\citeauthoryear{{Springel}}{{Springel}}{2005}]{Springel2005b}
{Springel} V.,  2005, \mn@doi [MNRAS] {10.1111/j.1365-2966.2005.09655.x}, \href
  {http://adsabs.harvard.edu/abs/2005MNRAS.364.1105S} {364, 1105}

\bibitem[\protect\citeauthoryear{{Springel}}{{Springel}}{2010}]{Arepo}
{Springel} V.,  2010, \mn@doi [\mnras] {10.1111/j.1365-2966.2009.15715.x},
  \href {http://adsabs.harvard.edu/abs/2010MNRAS.401..791S} {401, 791}

\bibitem[\protect\citeauthoryear{{Springel}}{{Springel}}{2011}]{Springel2011a}
{Springel} V.,  2011, preprint, \href
  {http://adsabs.harvard.edu/abs/2011arXiv1109.2218S} {} (\mn@eprint {arXiv}
  {1109.2218})

\bibitem[\protect\citeauthoryear{{Stinson}, {Brook}, {Macci{\`o}}, {Wadsley},
  {Quinn}  \& {Couchman}}{{Stinson} et~al.}{2013}]{Stinson2013}
{Stinson} G.~S.,  {Brook} C.,  {Macci{\`o}} A.~V.,  {Wadsley} J.,  {Quinn}
  T.~R.,   {Couchman} H.~M.~P.,  2013, \mn@doi [\mnras] {10.1093/mnras/sts028},
  \href {http://adsabs.harvard.edu/abs/2013MNRAS.428..129S} {428, 129}

\bibitem[\protect\citeauthoryear{{Stone}, {Gardiner}, {Teuben}, {Hawley}  \&
  {Simon}}{{Stone} et~al.}{2008}]{Athena}
{Stone} J.~M.,  {Gardiner} T.~A.,  {Teuben} P.,  {Hawley} J.~F.,   {Simon}
  J.~B.,  2008, \mn@doi [\apjs] {10.1086/588755}, \href
  {http://adsabs.harvard.edu/abs/2008ApJS..178..137S} {178, 137}

\bibitem[\protect\citeauthoryear{{Teyssier}}{{Teyssier}}{2002}]{Teyssier2002}
{Teyssier} R.,  2002, \mn@doi [\aap] {10.1051/0004-6361:20011817}, \href
  {http://adsabs.harvard.edu/abs/2002A%26A...385..337T} {385, 337}

\bibitem[\protect\citeauthoryear{Thomas \& Lombard}{Thomas \&
  Lombard}{1979}]{Thomas1979}
Thomas P.~D.,  Lombard C.~K.,  1979, \mn@doi [AIAA Journal] {10.2514/3.61273},
  17, 1030

\bibitem[\protect\citeauthoryear{Toro}{Toro}{1999}]{Toro}
Toro E.~F.,  1999, {Riemann Solvers and Numerical Methods for Fluid Dynamics: A
  Practical Introduction}, 2nd edn.
Springer, Berlin

\bibitem[\protect\citeauthoryear{{Vogelsberger}, {Sijacki}, {Kere{\v s}},
  {Springel}  \& {Hernquist}}{{Vogelsberger} et~al.}{2012}]{Vogelsberger2012}
{Vogelsberger} M.,  {Sijacki} D.,  {Kere{\v s}} D.,  {Springel} V.,
  {Hernquist} L.,  2012, \mn@doi [\mnras] {10.1111/j.1365-2966.2012.21590.x},
  \href {http://adsabs.harvard.edu/abs/2012MNRAS.425.3024V} {425, 3024}

\bibitem[\protect\citeauthoryear{{Vogelsberger} et~al.,}{{Vogelsberger}
  et~al.}{2014}]{Vogelsberger2014Illustris}
{Vogelsberger} M.,  et~al., 2014, \mn@doi [\nat] {10.1038/nature13316}, \href
  {http://adsabs.harvard.edu/abs/2014Natur.509..177V} {509, 177}

\bibitem[\protect\citeauthoryear{{Wadsley}, {Stadel}  \& {Quinn}}{{Wadsley}
  et~al.}{2004}]{Wadsley2004}
{Wadsley} J.~W.,  {Stadel} J.,   {Quinn} T.,  2004, \mn@doi [\na]
  {10.1016/j.newast.2003.08.004}, \href
  {http://adsabs.harvard.edu/abs/2004NewA....9..137W} {9, 137}

\bibitem[\protect\citeauthoryear{{Yee}, {Vinokur}  \& {Djomehri}}{{Yee}
  et~al.}{2000}]{Yee2000a}
{Yee} H.~C.,  {Vinokur} M.,   {Djomehri} M.~J.,  2000, \mn@doi [Journal of
  Computational Physics] {10.1006/jcph.2000.6517}, \href
  {http://adsabs.harvard.edu/abs/2000JCoPh.162...33Y} {162, 33}

\bibitem[\protect\citeauthoryear{{Zhu}, {Chang}, {van Kerkwijk}  \&
  {Wadsley}}{{Zhu} et~al.}{2013}]{Zhu2013}
{Zhu} C.,  {Chang} P.,  {van Kerkwijk} M.~H.,   {Wadsley} J.,  2013, \mn@doi
  [\apj] {10.1088/0004-637X/767/2/164}, \href
  {http://adsabs.harvard.edu/abs/2013ApJ...767..164Z} {767, 164}

\bibitem[\protect\citeauthoryear{{Zhu}, {Hernquist}  \& {Li}}{{Zhu}
  et~al.}{2015}]{Zhu2015}
{Zhu} Q.,  {Hernquist} L.,   {Li} Y.,  2015, \mn@doi [\apj]
  {10.1088/0004-637X/800/1/6}, \href
  {http://adsabs.harvard.edu/abs/2015ApJ...800....6Z} {800, 6}

\bibitem[\protect\citeauthoryear{de Vasconcellos \& da Silva}{de~Vasconcellos
  \& da~Silva}{2014}]{Vasconcellos2014}
de Vasconcellos J.,  da Silva D.,  2014, \mn@doi [Journal of the Brazilian
  Society of Mechanical Sciences and Engineering] {10.1007/s40430-014-0223-2},
  pp~1--9

\bibitem[\protect\citeauthoryear{van Leer}{van Leer}{1984}]{vanLeer1984}
van Leer B.,  1984, \mn@doi [SIAM Journal on Scientific and Statistical
  Computing] {10.1137/0905001}, 5, 1

\makeatother
\end{thebibliography}

\appendix

\section{Yee Vortex}
\label{app:yee}
For the setup of the Yee vortex we follow \citet{Yee2000a}. The mesh
is defined on the domain $\left[-5,5\right] \times
\left[-5,5\right]$. In the initial conditions, we set density $\rho$,
velocity $\textbf{v}$, and specific internal energy $u$ of the cells
to the value of the continuous fields at the centres of mass of the
cells. The continuous fields at position $\textbf{r} =
\left(x,y\right)$ are given by

 \begin{eqnarray*}
   T \left( \textbf{r} \right) &=& T_{\inf} - \frac{ \left( \gamma-1
                                   \right)  \beta^2 }{ 8 \gamma \pi^2
                                   } e^{1-r^2} , \\
   \rho \left( \textbf{r} \right) &=& T^{\frac{1}{1-\gamma}} , \\
   v_x \left( \textbf{r} \right) &=& -y \frac{ \beta }{ 2\pi } e^{
                                     \frac{1-r^2}{2} } , \\
   v_y \left( \textbf{r} \right) &=&  x \frac{ \beta }{ 2\pi } e^{
                                     \frac{1-r^2}{2} } , \\
   u \left( \textbf{r} \right) &=& \frac{T}{\gamma-1}.
 \end{eqnarray*}
We choose the parameters as $T_{\inf} = 1$, $\gamma = 1.4$, and $\beta = 5$.
 
The setup of a Cartesian mesh is trivial. To organise the
mesh-generating points on rings around the center of the vortex in a
regular way, we first compute the width of a ring as
$d_{\mathrm{ring}} = 10 / N$, where $10$ is the size of our domain in
one dimension and $N$ the linear resolution. We then add
mesh-generating points on circles with radii that are multiples of the
width of a ring, $r_{\mathrm{ring}} = i \times d_{\mathrm{ring}}$. On
every circle we equidistantly add $2 \pi \, r_{\mathrm{ring}} /
d_{\mathrm{ring}}$ points. We only add points that lie in the
computational domain and repeat this until none of the points of a new
ring are in the domain anymore.

\section{Keplerian Disc}
\label{app:disk}

The setup for the Keplerian disc is similar to the one used in
\citet{HopkinsGizmo}. We use a computational domain of
$\left[-2.5,2.5\right] \times \left[-2.5,2.5\right]$.  The disc is
cold, i.e.~the pressure support is negligible compared to
gravitational force and orbital velocity. The initial mesh is again
organised on rings around the center of the disk. We set the initial surface
density $\rho$, velocity $\textbf{v}$, and specific internal energy
$u$ of the cells to the values of the continuous fields at the centres
of mass of the cells. The continuous fields at position $\textbf{r} =
\left(x,y\right)$ and $r \equiv \left| r \right|$ are given as
follows.  For $r < 0.5$ and $r > 2$ we set
\begin{eqnarray*}
   \rho &=& 10^{-5} , \\
   \textbf{v} &=& 0 , \\
   u &=& \frac{ 5 \gamma }{ 2 \rho } \times 10^{-5}.
\end{eqnarray*}
And for $0.5 \leq r \leq 2$ we adopt
\begin{eqnarray*}
   \rho &=& 1.0 , \\
   v_x &=& - \frac{y}{ r } \sqrt{ \frac{1}{r} } , \\
   v_y &=& \frac{x}{ r } \sqrt{ \frac{1}{r} } , \\
   u &=& \frac{ 5 \gamma }{ 2 \rho } \times 10^{-5}.
\end{eqnarray*}
We choose an adiabatic index of $\gamma = 5/3$.  The constant external
gravitational acceleration is given by
\begin{equation}
  \textbf{g} = - \frac{ \textbf{r} }{ r \left(  r^2 + \epsilon^2 \right) },
\end{equation}
where $\epsilon = 0.25$ for $r < 0.25$ and $\epsilon = 0$
everywhere else.

\bsp
\label{lastpage}

\end{document}